%% file: arxiv_v1.tex
\documentclass[10pt,conference,letterpaper]{IEEEtran}
\usepackage{times,amsmath,epsfig}

\usepackage{booktabs} 
\usepackage{epigraph}
\usepackage{multirow}

\usepackage{framed}
\usepackage{multirow}
\usepackage{graphics}
\usepackage{url}
\usepackage{wrapfig}
\usepackage{algorithmic} 
\usepackage{algorithm} 
\usepackage{float} 
\usepackage{amsmath}
\usepackage{amssymb}
\usepackage{xspace}
\usepackage{amsfonts}
\usepackage{xtab,afterpage}
\usepackage{comment}
\title{Managing App Install Ad Campaigns in RTB: A Q-Learning Approach}
\author{
{Anit~Kumar~Sahu{\small $~^{\#1}$}, Shaunak~Mishra{\small $~^{*2}$}, Narayan~Bhamidipati{\small $~^{*3}$} }%
\vspace{1.6mm}\\
\fontsize{10}{10}\selectfont\itshape
$^{\#}$\,Carnegie Mellon University, Pittsburgh, USA\\
\fontsize{9}{9}\selectfont\ttfamily\upshape
$^{1}$\,anits@andrew.cmu.edu
\vspace{1.2mm}\\
\fontsize{10}{10}\selectfont\rmfamily\itshape
$^{*}$\,Yahoo Research, Sunnyvale, USA\\
\fontsize{9}{9}\selectfont\ttfamily\upshape
\{$^{2}$\,shaunakm, $^{3}$\,narayanb\}@yahoo-inc.com
}
\begin{document}
\maketitle
\begin{abstract}
Real time bidding (RTB) enables demand side platforms (bidders) to scale ad campaigns across multiple publishers affiliated to an RTB ad exchange.
While driving multiple campaigns for mobile app install ads via RTB, the bidder typically has to: (i) maintain each campaign's efficiency (\emph{i.e.}, meet advertiser's target cost-per-install),
(ii) be sensitive to advertiser's budget, and (iii) make profit after payouts to the ad exchange. In this process,
there is a sense of delayed rewards for the bidder's actions; the exchange charges the bidder right after the ad is shown,
but the bidder gets to know about resultant installs after considerable delay.
This makes it challenging for the bidder to decide beforehand the bid (and corresponding cost charged to advertiser) for each ad display opportunity.
To jointly handle the objectives mentioned above,
we propose a state space based policy which decides the exchange bid and advertiser cost for each opportunity.
The state space captures the current efficiency, budget utilization and profit.
The policy based on this state space is trained on past decisions and outcomes via a novel Q-learning algorithm which accounts for the delay in install notifications.
In our experiments based on data from app install campaigns managed by Yahoo's Gemini advertising platform,
the Q-learning based policy led to a significant increase in the profit and number of efficient campaigns.
\end{abstract}

\input{introduction.tex}
\input{setup.tex}
\input{state_space.tex}
\input{q_learning.tex}
\input{results.tex}
\bibliographystyle{IEEEtran}
\bibliography{ICDE_CIKM}
\input{appendix.tex}

\end{document}

%% file: introduction.tex
\section{Introduction} \label{sec:introduction}
In 2017, the revenue generated from ads for mobile apps was more than $33$ billion USD \cite{install_ad_spending_2017}.
This is not surprising given that there are over 5 million apps (in Google PlayStore and Apple App Store) \cite{num_apps}, 
and most apps struggle to achieve a large user base. To attract more users, apps naturally resort to online advertising platforms.
Such platforms (\emph{e.g.}, Yahoo's Gemini \cite{gemini_mappi}), drive app install campaigns by showing ads on owned-and-operated properties
(\emph{e.g.}, Yahoo mail, Tumblr and Yahoo Finance in case of Gemini) as well as third party publishers via an external RTB ad exchange (\emph{e.g.}, MoPub).
Such RTB ad exchanges offer diversity and scale for app install advertisers; but they also introduce new challenges as described below. 

\begin{figure}[h]
\centering
  \includegraphics[width=0.9\columnwidth]{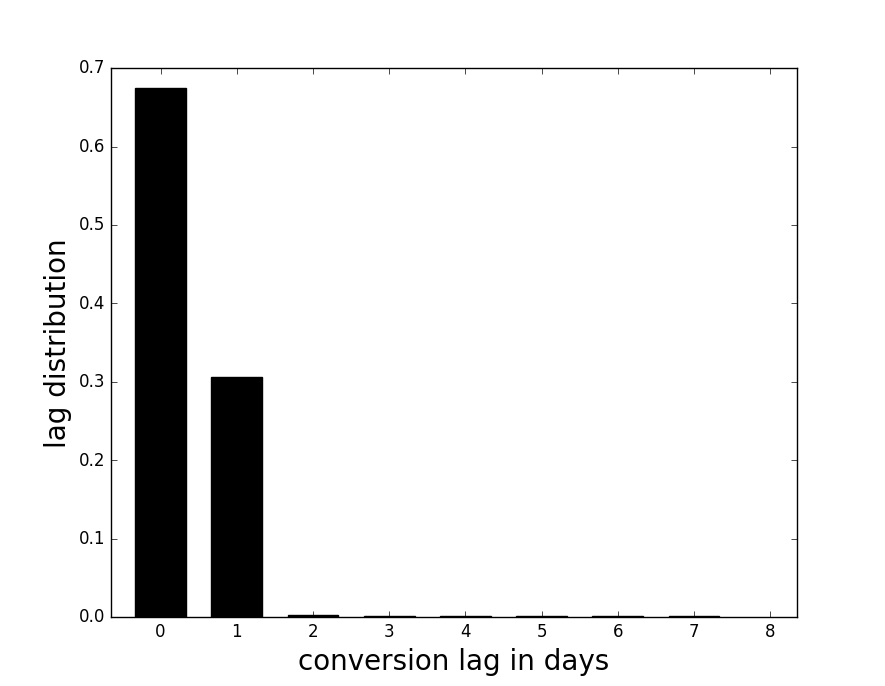}
  \caption{Conversion delay distribution for app installs, \emph{i.e.}, the time between click and first use of the app (for app install ads managed by Yahoo Gemini).
Majority of the conversions are received within $2$ days of the click.}
  \label{fig:conversion_lag}
\end{figure}

In an RTB ad exchange, multiple bidders participate in an auction for each ad display request issued by a publisher (via the exchange).
Each bidder could be managing multiple campaigns at the same time,
and showing ads across multiple publishers through the exchange.
In practice, there is considerable heterogeneity across publishers in terms of ad request volume, audience quality and auction floor prices.
The bidder's profit also keeps evolving over time, and so do the efficiencies across campaigns (depending on costs charged to advertisers).
So when a bidder selects a campaign for an ad request, while deciding how much should it bid and how much should it charge the advertiser,
it naturally faces the question: is my bid and cost for the campaign worth it given the publisher, my current profit and the campaign's current efficiency?
Sensitivity to advertiser's budget adds yet another  constraint for the bidder.
Intuitively, feedback from past decisions and outcomes can assist the bidder in coming up with
better future decisions amidst the challenges mentioned above. However, unlike regular cost-per-click (CPC) ads \cite{broder2008computational} where feedback (click/no-click) is fast, in the case of app install ads there can be considerable delays in knowing whether the user installed the app after clicking on the ad.
The delay stems from the following issue: after clicking an app install ad,
the user is typically taken to the Google PlayStore or Apple App Store (external to the bidder, publisher and advertiser).
The bidder gets to know about the ad conversion (install) only when the user opens the installed app for the first time (typically conveyed by the app advertiser or third parties).
Such delays can span days as shown in Figure~\ref{fig:conversion_lag} (for app install ads managed by Yahoo Gemini).

Current RTB literature focuses on just some of the above mentioned objectives.
In \cite{lin2016combining,fernandez2016optimal} the focus is purely on profit maximization;
whereas \cite{zhang2016feedback} focuses purely on campaign efficiency.
In the context of learning from past decisions and outcomes, \cite{cai2017real} employed value iteration (a form of reinforcement learning \cite{gosavi2009reinforcement}) to solely optimize for campaign efficiency for CPC ads. 
In fact, the notion of campaign efficiency (\emph{i.e.}, the discrepancy in cost-per-action delivered versus the advertiser's target cost-per-action) in current RTB literature is mostly oriented towards clicks as actions; this does not involve feedback delays as described above for app installs.
To the best of our knowledge, our work is the first to address both profit and campaign efficiency a coupled manner, specifically in the context of app install ads.
We develop a state space framework, and leverage Q-learning \cite{gosavi2009reinforcement} (also a form of reinforcement learning) to learn from the outcomes of past decisions.
Our main contributions can be summarized as listed below:
\begin{enumerate}
	\item a state space approach which encompasses campaign's efficiency, advertiser's budget and bidder's profit,
	\item a Q-learning algorithm to learn a state space based policy for determining the bid at the exchange and cost charged to advertiser for each ad request. The novelty lies in our design of the reward function which accounts for feedback delays.
\end{enumerate}
The remainder of the paper is organized as follows.
Section~\ref{sec:setup} covers the paper's setup, and problem formulation.
In Section~\ref{sec:state_space} we discuss our state state approach in Section~\ref{sec:q_learning} we explain the proposed Q-learning algorithm. Finally, in Section~\ref{sec:results}, we describe our experimental results based on mobile app install ads data from Yahoo Gemini \cite{gemini_mappi}.

%% file: setup.tex
\section{Setup} \label{sec:setup}
In this section,
we first provide some background on online advertising via RTB.
This is followed by a description of the exact setup considered in this paper and underlying objectives.
Our setup is fairly standard in the online
advertising industry \cite{broder2008computational},
and resembles the Yahoo Gemini offering for app install advertisers; the primary motivation behind this paper.
\subsection{Online advertising via RTB}
In RTB, several bidders participate at an ad exchange,
and bid for ad display opportunities provided by publishers affiliated to the exchange.
The typical sequence of events that takes place during an RTB auction can be described as follows.
When a user visits a publisher (\emph{e.g.}, website, app), the publisher conveys an ad display opportunity to the
exchange; this includes details like the user's identifier (\emph{e.g.}, mobile IDFA or AAID) and floor price for the auction.
The ad exchange then relays this information to the bidders. At this stage, an interested bidder finds the best matching ad (from its current list of campaigns).
Such a match might be based on the predicted click-through-rate (pCTR, \emph{i.e.}, $P(click | impression)$),
and predicted conversion/install rate (pCVR, \emph{i.e.}, $P(install | click)$) associated with the display opportunity; in particular,
details on ranking app install ads via pCTR and pCVR models can be found in \cite{mappi_CIKM}.
Having selected the best ad for the opportunity, the bidder decides: (i) its bid at the auction, and (ii) the corresponding cost to be charged to the advertiser.
If the bidder wins in the (second-price) auction \cite{broder2008computational}, it has to pay the exchange only if its ad is shown to the user (\emph{i.e.}, receives impression).
However, under the CPC pricing model \cite{broder2008computational}, the bidder can charge the advertiser (\emph{i.e.}, the determined cost) only if an user clicks on the shown ad.

Although an app install advertiser is charged for clicks, it is typically interested in campaign efficiency. Efficiency $\eta$ is defined as $\displaystyle{\frac{actual~CPI}{target~CPI}}$,
where $CPI$ stands for cost-per-install (\emph{i.e.}, the app-install equivalent of cost-per-action).
The $actual~CPI$ is the total cost charged to the advertiser divided by total installs received.
To capture a notion of satisfaction across all app-install advertisers associated with a bidder,
we define \textit{happy} campaigns as the count of campaigns with $\eta$ below $1+\epsilon$, where $\epsilon$
represents the tolerance relative to the target CPI provided by advertisers.
The differences in when the bidder is charged versus the advertiser is charged,
and the observation that the app-install advertiser is more concerned about CPI than CPC,
leave quite some room for the bidder to optimize its bid and cost decisions for each display opportunity.

\subsection{Problem formulation}
We consider a bidder which is handling $N$ app-install ad campaigns; campaign $i$ has a target $CPI^*_i$ and budget $B_i$ for time horizon $T$ (same across campaigns).
The (sole) ad exchange, with which the bidder interacts,
is associated with $M$ publishers. The bidder's cumulative spend on publisher $j$ at time $t$ is denoted by $spend_{j,t}$,
and represents the amount paid by the bidder to the exchange for the impressions shown on publisher $j$.
Similarly, we denote the cumulative advertiser cost (across all advertisers) charged for ads shown on publisher $j$ by $cost_{j,t}$.
The margin $m_{j,t}$ is defined as $\displaystyle{\frac{cost_{j,t} - spend_{j,t}}{\sum_{j=1}^M spend_{j,t}}}$, which is indicative of the relative profit/loss
being made on publisher $j$.
The bidder's goal is to maximize the number of happy campaigns (with $\eta_{i,T} < 1+\epsilon$ ), and the overall margin (\emph{i.e.},
$\sum_{j=1}^M m_{j,T}$) at the end of the time horizon, \emph{i.e.}, at $t=T$.

%% file: state_space.tex
\section{State space approach} \label{sec:state_space}
We first describe an intuitive approach (in Section~\ref{subsec:intuitive_approach})
which worked reasonably well in
our experiments. Drawing insights from this intuitive approach,
we then describe the detailed state space formulation in Section~\ref{subsec:state_space_formulation}; the proposed Q-learning algorithm (in Section~\ref{sec:q_learning}) is based on this state space.
\subsection{An intuitive approach}\label{subsec:intuitive_approach}
\begin{figure}[h]
\centering
  \includegraphics[width=0.99\columnwidth]{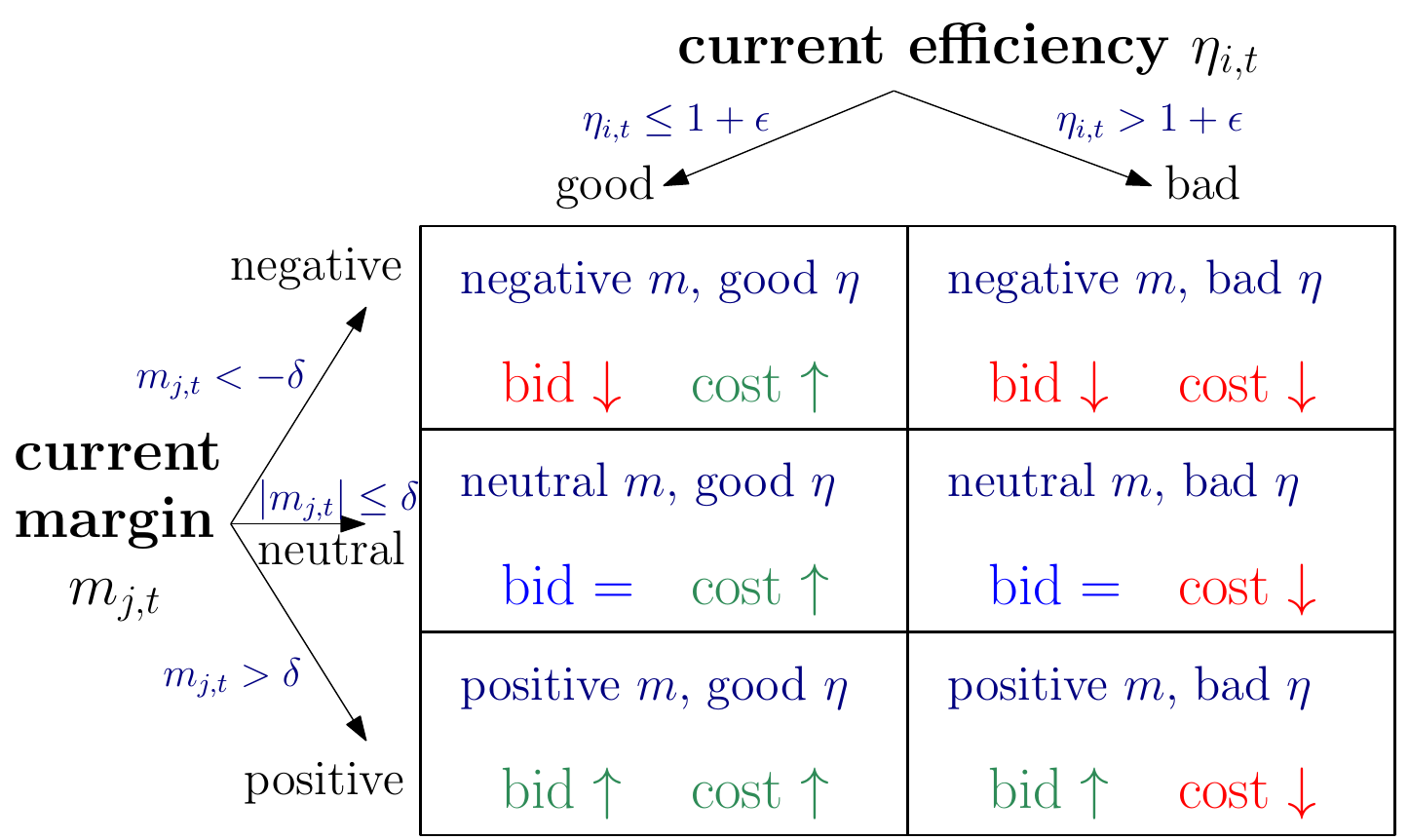}
  \caption{An intuitive approach for updating bid and cost based on current publisher margin $m_{j,t}$
and current campaign efficiency $\eta_{j,t}$. The symbols $\uparrow$, $\downarrow$ and $=$ stand for increase, decrease and no change respectively; $\delta,\epsilon>0$ represent the thresholds for good efficiency and positive margin.}
  \label{fig:intuitive_approach}
\end{figure}

Consider a point of time $t$, where the bidder is placing a bid for campaign $i$ for an opportunity in
publisher $j$.
Assume that the bidder had already computed the bid (at the exchange),
and the cost (to the advertiser) for this particular opportunity based on data from the past.
This could be the standard expected-cost-per-impression ($eCPM$) bid \cite{gemini_mappi,broder2008computational} based on the pCTR and pCVR associated
with the opportunity (\emph{i.e.}, bid $=$ eCPM $=$ Target CPI$\times$pCVR$\times$pCTR, and cost $=$ Target CPI$\times$pCVR).
But at this point of time, the bidder suddenly gets to know the current publisher margin ($m_{j,t}$),
and campaign efficiency ($\eta_{j,t}$), and has the option of update the bid and cost.
An intuitive approach to do so based on $m_{j,t}$ and $\eta_{j,t}$ (and also keeping in mind the goals outlined in Section~\ref{sec:setup})
would be as shown in Figure~\ref{fig:intuitive_approach}.
For example, when the current efficiency is bad and margin is negative, the bidder is better off reducing both its bid and cost. Increasing the bid would hurt the margin, while increasing the cost would hurt the efficiency. But when the margin is negative and efficiency is good, the bidder can afford to charge the advertiser more (by increasing the cost) while decreasing its bid for better margin.
At a high level, Figure~\ref{fig:intuitive_approach} defines a discrete state space
based on current efficiency and margin, and then takes an intuitive step based on the state.
In Section~\ref{subsec:state_space_formulation} we generalize such a state space approach, and later in Section~\ref{sec:q_learning}, we show how one can learn the best action (\emph{i.e.}, whether to decrease/increase and the magnitude of change) for a particular state. 
\subsection{State space formulation}\label{subsec:state_space_formulation}
Drawing motivation from the intuitive approach in Section~\ref{subsec:intuitive_approach},
we define a discrete state space $\mathcal{S} = \mathcal{S}_m \times \mathcal{S}_\eta \times \mathcal{S}_B$, where 
$\mathcal{S}_m$, $\mathcal{S}_\eta$, and $\mathcal{S}_B$ represent quantized publisher margin, campaign efficiency, and campaign budget respectively. The main features of such a state space are described below.

\subsubsection{Quantization}
Quantization is done on the basis of domain knowledge.
For margin, we consider a bin around zero of the form $(-\Delta_m, \Delta_m]$, and $l_m$ uniformly sized bins (width $2\Delta_m$) to the left and to the right of the 'zero' bin (similar to the quantization in Figure~\ref{fig:intuitive_approach}).
For efficiency, we partition the intervals $(1+\epsilon,\eta_{upper}]$ and $[0, 1+ \epsilon]$ into $l_\eta$ uniform bins each (where $\eta_{upper}$ is a predetermined upper bound).
For budget, the fraction of budget remaining is binned into $2l_B$ uniformly spaced bins in the interval $[0,1]$.
As a result of the above quantizations, we obtain a discrete state space which not only captures our objectives,
but also simplifies the policy learning process (as described in Section~\ref{sec:q_learning}).
Also, the granularity of the state space is such that it is not expensive to infer the current state of the system;
in a large scale setup with thousands of campaigns and publishers, maintaining near real time aggregates of publisher wise margins and campaign wise costs is way simpler than
aggregates for each (publisher, campaign) pair.

\subsubsection{Controllability}
A natural question that arises in any state space based dynamical system is if it is possible
to drive the system from any initial state to any final state via a (finite) sequence of inputs (\emph{i.e.}, controllability \cite{controls_book}). In our setup, at least two factors make the system inherently uncontrollable:
(i) finite advertiser budget,
and (ii) variability in the volume of ad requests from a publisher.
As a result, we cannot employ any generic control system which assumes controllability. 
However, some states are reachable from any initial state, and hence our setup satisfies a weaker version
of controllability called reachability \cite{controls_book}.
The challenging part in our setup is to drive a (publisher, campaign) pair to a desirable state to eventually meet our objectives.

\subsubsection{State based bid and cost updates}\label{subsubsec:update_forms}
To drive the current state of a (publisher, campaign) pair to desirable states,
a simple strategy is to have additive updates to the bid and cost depending on the current state.
In this paper, we consider
additive bid and cost updates of the following form:
	\begin{align} \label{eq:update_forms}
	bid_{i,j,t} &= bid_{i,j,t}  + \kappa_{bid} f_m\left(\hat{m}_{j,t}\right) bid_{i,j,t}, \\
	cost_{i,j,t} &= cost_{i,j,t} + \nonumber \\ 
         & \quad \;  \left(1+\kappa_{\beta} f_\beta \left(\hat{\beta}_{i,t}\right) \right)\kappa_{\eta}  f_\eta \left(\hat{\eta}_{i,t}\right)cost_{i,j,t},
	\end{align}
where $\kappa_{bid}$, $\kappa_{\beta}$, and $\kappa_{\eta}$ are constants.
Discrete functions $f_{m}(\cdot)$, $f_{\beta}(\cdot)$, $f_{\eta}(\cdot)$ map the quantized versions of current margin ($\hat{m}_{j,t}$),
leftover budget fraction ($\hat{\beta}_{i,t}$),
and current efficiency ($\hat{\eta}_{i,t}$) to discrete values.
For further simplification, we assume that: (i) $f_\beta (\hat{\beta}_{i,t}) = \hat{\beta}_{i,t}$,
and (ii) the range (\emph{i..e}, the discrete set of possible output values)
of $f_{m}(\cdot)$ and $f_{\eta}(\cdot)$ is fixed for our setup (determined using domain knowledge).
For example, the set of possible output values for $f_{m}(\cdot)$
could be $\{-1, -0.5, 0, 0.5, 1\}$. Hence, the remaining task at hand is basically to 'learn'
which state should map to which output value for functions $f_{m}(\cdot)$ and $f_{\eta}(\cdot)$, so that our end goals are met;
this is precisely what we cover in Section~\ref{sec:q_learning}.
For consistency with standard methods in the online advertising industry  \cite{gemini_mappi,broder2008computational,mappi_CIKM},
we assume that, before each update, the bid equals eCPM ($=$ Target CPI$\times$pCVR$\times$pCTR)
and cost equals Target CPI$\times$pCVR; note that the pCVR and pCTR can vary for each display opportunity due to differences in features derived from the campaign, publisher and online user \cite{mappi_CIKM}.

%% file: q_learning.tex
\section{Q-learning}  \label{sec:q_learning}
In general,
there are two ways one can go about learning functions $f_{m}(\cdot)$, and $f_{\eta}(\cdot)$
described in Section~\ref{subsubsec:update_forms}. One way is to mimic the entire RTB setup via a state transition model,
and optimize $f_{m}(\cdot)$, and $f_{\eta}(\cdot)$ around it.
But building such a complex model is not practically feasible,
and the learnt $f_{m}(\cdot)$, and $f_{\eta}(\cdot)$
would seriously suffer from modelling errors.
Another way is a model free approach, \emph{e.g.}, Q-learning which learns $f_{m}(\cdot)$, and $f_{\eta}(\cdot)$ directly from
past decisions and outcomes.
In Section~\ref{subsec:qlsetup},
we describe the proposed Q-learning
algorithm for our setup; this is followed by a detailed description of the associated Q-learning reward function in Section~\ref{subsec:reward}.
\subsection{Q-learning}
\label{subsec:qlsetup}
As mentioned in Section~\ref{subsubsec:update_forms}, at each update step, we take a compound action (in the form of changes in bid and cost).
Due to the discrete nature of the functions $f_{m}(\cdot)$, and $f_{\eta}(\cdot)$, there are only a finite number of actions that can be taken in each state.
Thus, if there are $a_m$ possible values for $f_{m}(\cdot)$, and $a_{\eta}$ possible values for $f_{\eta}(\cdot)$, there are $a_m\times a_\eta$ possible
actions $a \in \mathcal{A}$ (\emph{i.e.}, the action space which is the product of  ) for any state $s \in  \mathcal{S} = \mathcal{S}_m \times \mathcal{S}_\eta \times \mathcal{S}_B$.
The standard Q-learning update step \cite{gosavi2009reinforcement} can now be stated in our context as follows:
\begin{align}
Q\left(s_{t},a_{t}\right) &= \left(1-\alpha\right)Q\left(s_{t},a_{t}\right)\nonumber\\& \quad +\alpha\left(R\left( t \right)+\gamma\max_{a \in \mathcal{A}}Q\left(s_{t+1},a\right)\right), \label{eq:q_update}
\end{align}
where $s_{t}$ and $a_{t}$ represent the state and action at time $t$, $R_{t}$ is the reward at time $t$, $\alpha_{t}$ is the learning rate, and $\gamma$ is the forgetting factor, and $\mathcal{A}$ is the space of possible actions.
At each time, the bidder selects an action, and observes the corresponding reward (to be defined in Section~\ref{subsec:reward}), and then updates the $Q$ value.
We describe below some important properties associated with this update in our context.

\subsubsection{Learning rate and forgetting factor} The learning rate $\alpha_{t}$ determines to what extent the newly acquired information
is weighed in comparison to the old information.
Theoretically, a decaying $\alpha_{t}$ ensures convergence, but results in very slow convergence rates, hence a small but constant value of $\alpha_{t}$ suffices.
The forgetting factor $\gamma \in [0,1]$ determines the importance of incorporation of future rewards. A low $\gamma$ makes the agent myopic as it then considers only current rewards, while $\gamma$ close to $1$ makes the agent strive for long-term high rewards.

\subsubsection{Exploration vs. exploitation} Note that the maximization step in \eqref{eq:q_update}, \emph{i.e.}, $\displaystyle{\max_{a \in \mathcal{A}}Q\left(s_{t+1},a\right)}$ is a greedy procedure.
The convergence of the algorithm depends on the balance between exploration and exploitation.
To converge faster, a natural step is to resort to an $\epsilon$-greedy policy, where with probability $1-\epsilon$ one chooses the 'max' action in \eqref{eq:q_update},
and with probability $\epsilon$ one chooses a random action from the action space $\mathcal{A}$.
In particular, we resort to Boltzmann sampling \cite{gosavi2009reinforcement} in our setup.
This means that, during exploration, the probability of selection an action given a state, is given by $\displaystyle{p\left(a|s\right)=\frac{\exp\left(Q\left(s,a\right)/\theta \right)}{\sum_{a'}\exp\left(Q\left(s,a'\right)/\theta\right)}}$,
where the temperature parameter $\theta$ ,
is decayed slowly over time so as to slowly reduce the exploration.
In addition, the learning rate $\alpha$ can be chosen in a systematic way for different states,
so as to quicken the learning pace for states which are not visited often.
\subsubsection{Deterministic vs. stochastic policy} It is crucial to learn a deterministic policy for our setup,
as a stochastic policy might result in states drifting off from the actual objective.

Once the Q-learning algorithm converges,
the best action corresponding to a given state is given by $\arg\max_{a \in \mathcal{A} }Q\left(s, a\right)$.
\subsection{Q-learning: Reward Functions}
\label{subsec:reward}
Given the state space example in Figure~\ref{fig:intuitive_approach}, the following challenges are encountered while designing a suitable reward function for Q-learning.
\subsubsection{Unobservable spend and cost}
Keeping in mind a large scale setup with many publishers and campaigns,
we assume that the bidder maintains data only at a publisher and campaign granularity.  
This means, the bidder tracks only publisher-wise spend across all campaigns, 
and campaign-wise cost and installs across all publishers.
The bidder does not track publisher-wise spend at a campaign-level, or the advertiser-wise cost and installs at a publisher level.
This brings in some sense of \textit{unobservability} regarding the effectiveness of cost and bid update actions in our setup.
For instance, the margin of a publisher is affected by all bid and cost update actions undertaken for all campaigns
associated with the publisher.
Similarly, the efficiency of a campaign is affected by all bid and cost update actions involving publishers where the campaign is being bid for.
Hence, a suitable reward function should be able to attribute margin and efficiency changes of a publisher and campaign respectively to individual actions.

\subsubsection{Sparse and delayed rewards}
The transition of a publisher's margin from a negative margin state to a neutral/positive margin state in the interval of two consecutive actions is usually very unlikely. Similarly, the efficiency of a campaign is unlikely to change in the interval of two consecutive actions. The transition is brought about by a sequence of actions,
rather than just an action. Hence, a suitable reward function should reward actions through intermediate rewards rather than only on change of state, \emph{i.e.}, it should reward the change in margin and efficiency of a publisher and campaign respectively.

Keeping the above points in mind, we propose a reward function of the following form:
\begin{align}
\label{eq:reward_function}
&R\left(m_{j,t},\eta_{i,t},\mbox{bid}_{i,j,t},\mbox{cost}_{i,j,t}\right)\nonumber\\ &= (1-\lambda) \kappa_{rm_{j},i,t}\left(m_{j,t}-m_{j,t-1}\right)
+  \lambda \kappa_{r\eta_{i},j,t }\left(\eta_{i,t-1}-\eta_{i,t}\right),
\end{align}
where $\kappa_{rm_{j},i,t}$ and $\kappa_{r\eta_{i},j,t}$ are weights which map the amount of attribution to be assigned to the compound actions $\left(\mbox{bid}_{i,j,t},\mbox{cost}_{i,j,t}\right)$, and hyper-parameter $\lambda \in [0,1]$ trades-off the importance given to change in margin versus the change in efficiency.
In particular,  $\displaystyle{\kappa_{rm_{j},i,t} = \frac{B_{i}-cost_{i,t}}{\sum_{l=1}^{N}B_{l}-cost_{l,t}}}$,
and $\displaystyle{\kappa_{r\eta_{i},j,t} = \frac{\mbox{spend}_{j,t}}{\sum_{l=1}^{M}\mbox{spend}_{l,t}}}$.
The weight $\kappa_{r\eta_{i},j,t}$ incorporates the spend ratio of the publisher $j$ as compared to the total spend, into the reward, and hence tries to approximately attribute the cause of change of $\eta_{i,t-1}$ to $\eta_{i,t}$ to the compound action local to publisher $j$ and campaign $i$.
Similarly, $\kappa_{rm_{j},i,t}$ incorporates the ratio of the budget of campaign $i$ as compared to the total unutilized budget, into the reward and hence tries to approximately attribute the cause of change of $m_{j,t-1}$ to $m_{j,t}$ to the compound action local to publisher $j$ and campaign $i$.
The reward function proposed in \eqref{eq:reward_function} not only considers both the objectives of our setup, but also rewards actions at each step.

%% file: results.tex
\section{Results} \label{sec:results}
In this section, we discuss experimental results based on data from mobile app install campaigns managed by Yahoo Gemini \cite{gemini_mappi}.
The state space was quantized as follows: $4$ bins for efficiency with efficiency threshold $\epsilon=0.2$,
$3$ bins for margin, $4$ bins for left-over budget. The cardinality of range of functions $f_m(\cdot)$ and $f_\eta(\cdot)$ was
set to $8$ and $14$.
We considered a sample of $183$ publishers and $400$ campaigns.
Our training data covered one week of impression level data spanning the selected publishers and campaigns, and the testing data covered the following week. The policy evaluations were carried out using the RTB simulator that we describe in
Appendix~\ref{appendix:simulator}.
The baseline for the performance improvements stated below is a PI controller \cite{zhang2016feedback} just optimizing for the margin.

\begin{figure}[h]
\centering
  \includegraphics[width=1.03\columnwidth]{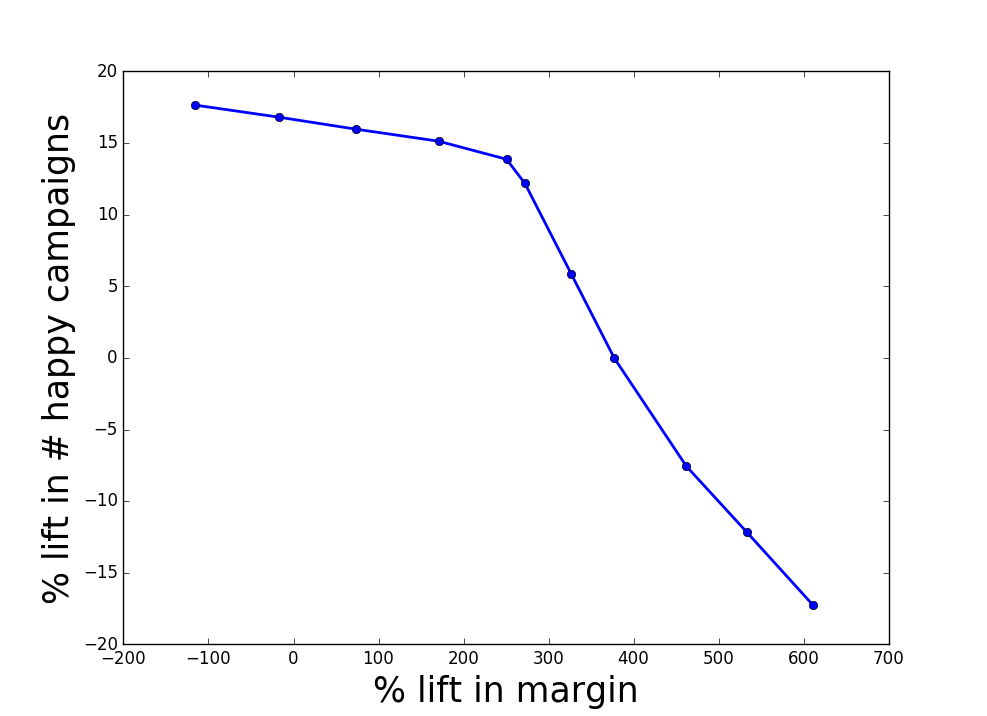}
  \caption{Trade-off between optimizing for margin versus optimizing for number of happy campaigns. As shown, after a sweet spot on the trade-off curve, the efficiency starts falling drastically.}
  \label{fig:tradeoff}
\end{figure}

\paragraph*{Margin-efficiency tradeoff}
As shown in \eqref{eq:reward_function}, the hyper-parameter $\lambda$ in the reward function provides a way to trade-off performance lifts in margin versus the number of happy campaigns.
Table~\ref{tab:tradeoff}, clearly shows this performance trade-off (as well the budget utilization) for different values of $\lambda$.
Note that $\lambda=0$ corresponds to a policy which focuses only on margin improvements, while at the other extreme,
$\lambda=1$ corresponds to a policy focused on efficiency improvement.
\begin{table}[!h]
	\centering
\begin{tabular}{|c|c|c|c|c|}
\toprule\textbf{$\lambda$}&\textbf{$\Delta$spend} $\%$&\textbf{$\Delta$margin} $\%$&\textbf{$\Delta$budget util.} $\%$ & \textbf{$\Delta$happy} $\%$\\ 
\midrule
$0$	& $-7.65$	       & $610.43$ & $71.57$ & $-17.23$ \\
\midrule
$0.1$ & $-4.01$      & $533.38$ & $67.05$ & $-12.18$  \\
\midrule
$0.2$ & $-0.07$     & $461.97$ & $63.99$ & $-7.56$ \\
\midrule
$0.3$ & $4.72$      & $377.06$	  & $59.52$ & $0.00$ \\
\midrule
$0.4$  & $7.83$     & $326.42$ & $56.68$ & $5.88$ \\
\midrule
$0.5$  &	$10.98$ &$272.07$	& $53.92$ &$12.18$ \\
\midrule
$0.6$ &      $11.52$ &$250.53$ & $50.29$& $13.87$ \\
\midrule
$0.7$ &	$12.14$ & $170.65$ & $38.70$	& $15.13$ \\
\midrule
$0.8$ &	$12.81$ & $73.25$ & $24.28$ & $15.97$ \\
\midrule
$0.9$ &	$13.37$ & $-17.58$ & $10.61$ & $16.81$ \\
\midrule
$1$	  & $14.05	$   &  $-115.83$ & $-4.27$ & $17.65$ \\
 \bottomrule\end{tabular}
	\caption{Variation in $\%$ lifts in overall spend, overall margin, overall budget utilization, and number of happy campaigns for different values of $\lambda$. 
}
	\label{tab:tradeoff}
\end{table}
Figure~\ref{fig:tradeoff} shows the trade-off between lift in margin versus the lift in number of happy campaigns; it also shows there is a \textit{sweet}
spot on the curve where a bidder might like to operate (leading to good margin without much efficiency loss).

\paragraph*{Discussion}
In our experiments, we observed that driving the margin and efficiency of a (publisher,campaign) pair to a favorable state becomes relatively easier with higher number of publishers and campaigns. It is interesting to note that the campaigns which start off as being unhappy have a higher chance of turning happy if bid for in multiple different publishers.
Campaigns which are bidded in only a few publishers, and are unhappy to start with, are less likely to be turned to happy campaigns.
The above observation leads to the question: how many publishers does a campaign need to be bid in, so as to be controllable, \emph{i.e.},
the efficiency of that campaign can be driven from any state to any desired state? This is indeed a question that needs further exploration and is of practical interest. We believe our
state space approach provides a reasonable framework for pursuing such questions, and our Q-learning approach is able to capture the performance trade-offs (\emph{e.g.}, efficiency vs. margin maximization) inherent to the setup.

%% file: appendix.tex
\appendix
\subsection{RTB simulator} \label{appendix:simulator}
A crucial part of the experiment section is a policy evaluator which emulates the RTB platform closely. Next, we describe the policy evaluator which consists of an RTB simulator. The details of the RTB simulator are as follows.
\begin{itemize}
	\item[$$]\emph{Auction granularity level}: We keep the auction granularity level to one minute, \emph{i.e.}, we assume that the bidding process takes place once in a minute. For each publisher, the bidder seeks the campaign with the maximum eCPM bid. For the next minute, we assume that for a publisher, the campaign with the highest eCPM bid is selected by the bidder for the entire minute.
	\item[$$] \emph{Bid and cost update}: We update the bids and costs of a particular campaign in a publisher, once in an hour and we update only those campaigns which won the auction in the last hour. To be particular, the compound action of updating the bids and costs of campaigns across different publishers is taken once in an hour.
	\item[$$] \emph{Campaign selection}: During the auction process, we only pick campaigns which have at least $10$ installs in our data set, so as to ensure we use campaigns which have fairly well estimated pCTR and pCVR values.
	\item[$$] \emph{Bid landscape}: The bidding landscape for each section is simulated as a sigmoid function of the form:
	\begin{align*}
	\mathbb{P}\left(win|bid\right)=\frac{1}{1+ae^{floorprice-bid}},
	\end{align*}
	where $a$'s are drawn from an uniform distribution for each publisher at the beginning of the simulator. If the bid is lower than the floor price, the win rate is allocated to be zero. The sigmoid function abstracts the chances of higher win rates with increasing bids. The floorprice is extracted from historical data for each publisher.
	\item[$$] \emph{pCTR and pCVR values}: We assume that we have a good estimate of pCTR and pCVR values for each (publisher,campaign) instance, where the pCTR and pCVR data is collected from historical data as the average pCTR and pCVR values over a period of one week.
	\item[$$] \emph{Budget Consideration}: We assign budgets to each campaign based on the past advertiser cost of each campaign. The advertiser cost of a campaign is a conservative estimate of the budget. However, it ensures, the budget estimate of a campaign is realistic as advertiser cost is less than the budget for a campaign.
	\item[$$] \emph{Discrete Policy}: The policy learnt, and hence evaluated, is discrete at all levels. The policy is discrete in terms of efficiency, leftover budget and, margin.
Discrete policies are expected to work better with more discretization of the state space (at the cost of higher complexity).
	\end{itemize}